\documentclass[a4paper,12pt]{article}

\usepackage{amsmath,amssymb}
\usepackage{cite}

\begin{document}

\author{Nikolay A. Kudryashov\footnote{E-mail: kudryashov@mephi.ru}, Nadejda B. Loguinova}

\title{Be careful with the Exp-function method}

\date{Department of Applied Mathematics, \\ National  Research Nuclear University MEPhI,  \\ 31 Kashirskoe Shosse, 115409, Moscow, \\ Russian Federation}

\maketitle


\begin{abstract}

An application of the Exp-function method to search for exact solutions of nonlinear differential equations is analyzed. Typical mistakes of application of the Exp-function method are demonstrated. We show it is often required to simplify the exact solutions obtained.  Possibilities of the Exp-function method and other approaches in mathematical physics are discussed. The application of the singular manifold method  for finding exact solutions of the Fitzhugh - Nagumo equation is illustrated.  The modified simple equation method is introduced. This approach is used to look for exact solutions of the generalized Korteweg - de Vries equation.

\end{abstract}

\newcommand {\psps} {\frac{\psi'}{\psi}}
\newcommand {\Dpsps} {\left( \frac{\psi'}{\psi} \right)'}
\newcommand {\DDpsps} {\frac{\psi''}{\psi}}
\newcommand {\DDDpsps} {\frac{\psi'''}{\psi}}
\newcommand {\pspsPP} {\left( \frac{\psi'}{\psi} \right)^2}
\newcommand {\pspsPPP} {\left( \frac{\psi'}{\psi} \right)^3}
\newcommand {\pspsPPPP} {\left( \frac{\psi'}{\psi} \right)^4}
\newcommand {\pspsDD} {\left( \frac{\psi'}{\psi} \right)''}
\newcommand {\DpspsPP} {\left( \left( \frac{\psi'}{\psi} \right)' \right)^2}
\newcommand {\kz} {\frac{z}{2} \sqrt{4c-1}}
\newcommand {\kw} {kx + \omega t}

\section{Introduction}

He and Wu introduced  the so called Exp-function method in
\cite{He01} to search for exact solutions of nonlinear differential
equations. At present this method is very popular and we can find a
lot of publications  with applications of this method  in
journals \cite{He02, Wu01, Zhu01, Abdou01, Elwakil01, Elwakil02,
Zhang01, Ebaid01, Yusufoglu01, Ozis01, Chun01, Bekir01, Zhang02,
Soliman01}. We can read a number of ecstatic words  about
possibilities of this method. Let us present here some of them.

"The expression of the Exp-function method is more general than the
sinh-function and tanh-function, so we can find more general
solutions in Exp-function method" \cite{Yusufoglu01}.

"Exp-function method is easy, concise and an effective method to
implement to nonlinear evolution equations arising in mathematical
physics" \cite{Ozis01}.

"All applications verified that the Exp-function method is
straightforward, concise and effective in obtaining generalized
solitary solutions and periodic solutions of nonlinear evolution
equations. The main merits of this method over the other methods are
that it gives more general solutions with some free parameters"
\cite{Chun01}.

"The solution procedure of this method can be easily extended to
other kinds of nonlinear evolution equations" \cite{Soliman01}.

"The Exp-function method leads to not only generalized solitonary
solutions but also periodic solutions" \cite{Zhang02}.

"Our first interest is implementing the Exp-function method to
stress its power in handling nonlinear equations, so that one can
apply it to models of various types of nonlinearity" \cite{Bekir01}.

We hope we have enough raptures about the Exp-function method.

The aim of this paper is to analyze some recent papers with
application of the Exp-function method and to discuss the main
deficiencies of this method.

We would like also to point out the niche which this method can occupy
in comparison with other methods for finding exact
solutions of nonlinear differential equations.

The idea of the Exp-function method is very simple \cite{He02, Wu01,
Zhu01, Abdou01, Elwakil01, Elwakil02, Zhang01, Ebaid01, Yusufoglu01,
Ozis01, Chun01, Bekir01, Zhang02, Soliman01} but results are very
cumbersome. If we look at papers \cite{Abdou01, Elwakil01,
Elwakil02} we will note that we can not check the most part of these
exact solutions. As the consequence of application of the Exp -
function method we can usually obtain exact solutions which have to
be simplified. However sometimes it is not easy to make
simplifications taking into account cumbersome expressions. We are
going to demonstrate that the main deficiency of the application of
the Exp-function method is reducibility of the obtained exact
solutions.

The paper is organized as follows. In section 2 we analyze the
application of the Exp-function method to the Korteweg - de Vries -
Burgers equation by Soliman \cite{Soliman01}. We simplify exact
solutions obtained by author and demonstrate that all his
solutions are not new and were found before. In section 2 we
consider the application of the Exp-function method to the Riccati
equation by Zhang \cite{Zhang02} and point out that these solutions
are also simplified and these solutions are not new as well. In
section 4 we analyze the work by Bekir and Boz \cite{Bekir01} with
application of the Exp-function method to the Sharma - Tisso - Olver
equation and show that this exact solution can be simplified too.
Section 5 of this paper is devoted to discussion of the application
of the Exp-function method and other methods for finding exact
solutions of nonlinear differential equations. In section 6 we
introduce the modification of the simplest equation method and in
section 7 we demonstrate the application of the modified simplest
equation method to look for solution of the generalized Korteweg -
de Vries equation.

\section{Application of the Exp-function method to the
Korteveg - de Vries - Burgers equation by Soliman}

Let us analyze the application of the Exp-function method
to search for exact solutions of the Burgers - Korteweg - de Vries -
Burgers equation by Soliman \cite{Soliman01}
\begin{equation}
\label{A2.1}u_t+\varepsilon\,u\,u_x+\mu\,u_{xxx}-\nu\,u_{xx}=0
\end{equation}

Exact solutions of this equation  in the form of the solitary wave
at $\varepsilon=1$ were first found in work \cite{Kudryashov01}
using the singular manifold method \cite{Weiss01, Weiss02}. It takes
the form
\begin{equation}
\label{A2.2}u=C_0+\frac{6\,\nu^2}{25\,\mu}-\frac{3\,\nu^2}{25\,\mu}\,
\left(1+\tanh{\left\{\frac{\nu\,z}{10\,\mu}\right\}}\right)^2,
\qquad z=x-C_0\,t-x_0
\end{equation}
where $C_0$ and $x_0$ are arbitrary constants.

Solution \eqref{A2.2} can be transformed to the following form
\begin{equation}
\label{A2.3}u=C_0+\frac{6\,\nu^2}{25\,\mu}-\frac{12\,\nu^2}{25\,
\mu\left(1+e^{-\frac{2\,\nu\,z}{10\,\mu}}\right)^2}, \qquad
z=x-C_0\,t-x_0
\end{equation}

Soliman in \cite{Soliman01} obtained  four "new exact solutions" of
Eq. \eqref{A2.1} using the Exp-function method. The first exact
solution (solution (15) in \cite{Soliman01}) was written in the form

\begin{equation}\begin{gathered}
\label{A2.4}u=\frac{\frac{4}{25}\left(\frac{25\varepsilon\,\mu\,a_{-1}-3\,
\nu^2\,b_0^{2}}{\varepsilon\,\mu\,b_{0}^{2}}\right)
\,e^{\eta}+\frac{4\,a_{-1}}{b_{0}}+a_{-1}\,
e^{-\eta}}{e^{\eta}+b_0+\frac{b_{0}^{2}}{4}\,e^{-\eta}}, \\
\\
\eta=k\,(x+\,\alpha\,t), \qquad
\alpha=-\frac{2}{25}\left(\frac{50\,\varepsilon\,\mu\,a_{-1}-
3\,\nu^2\,b_{0}^{2}}{\mu\,b_{0}^{2}}\right)
\end{gathered}\end{equation}

Solution \eqref{A2.4} of work \cite{Soliman01} satisfies Eq.
\eqref{A2.1} in the case $k=\frac{\nu\,z}{10\,\mu}$. However exact
solution \eqref{A2.4}  can be transformed taking into account the
following set of equalities

\begin{equation}\begin{gathered}
\label{A2.5}
u=\frac{\frac{4}{25}\left(\frac{25\varepsilon\,\mu\,a_{-1}-3\,
\nu^2\,b_0^{2}}{\varepsilon\,\mu\,b_{0}^{2}}\right)
\,e^{\eta}+\frac{4\,a_{-1}}{b_{0}}+a_{-1}\,
e^{-\eta}}{e^{\eta}+b_0+\frac{b_{0}^{2}}{4}\,e^{-\eta}}=\\
\\
=\frac{4\,a_{-1}}{b_{0}^{2}}-\frac{12\,\nu^2}{25\,\varepsilon
\,\mu\,\left(1+\frac{b_0}{2}\,e^{-2\,\eta}\right)^2}
\end{gathered}\end{equation}

Substituting $\eta$ and $\alpha$ from \eqref{A2.4} into \eqref{A2.5}
and taking into consideration  $k=\frac{\nu\,z}{10\,\mu}$,
$\frac{4\,a_{-1}}{b_{0}^{2}}=C_0-\frac{6\,\nu^2}{25\,\mu}$,
$x_0=-\ln{\frac{b_0}{2}}$ and $\varepsilon=1$ we obtain solution
\eqref{A2.3}. One can see that the solution \eqref{A2.4} is
transformed to solution \eqref{A2.2} in the case $\varepsilon=1$ and
\eqref{A2.4} is not new solution.

It was also obtained  the second "new exact solution"  of the
Korteweg - de Vries - Burgers equation in work \cite{Soliman01} in
the form
\begin{equation}\begin{gathered}
\label{A2.6}
u=\frac{-\frac{3}{100}\left(\frac{16\nu^2\,b_{1}^{3}+225\,
\varepsilon\,\mu\,a_{-1}}{\varepsilon\,\mu\,b_{1}^{3}}\right)\,e^{2\,\eta}-
\frac{1}{100}\frac{-16\,\nu^2\,b_{1}^{3}+675\,\varepsilon\,\mu\,a_{-1}}
{\varepsilon\,\mu\,b_{1}^{2}}\,e^{\eta}+a_{-1}\,e^{-\eta}}
{e^{2\eta}+b_1\,e^{\eta}-\frac{4\,b_{1}^{3}}{27}\,e^{-\eta}},\\
\\
\eta=k\,x+k\,\lambda\,t, \qquad
\lambda=\frac{3}{100}\,\left(\frac{8\,\nu^2\,b_{1}^{3}+225\,\varepsilon\,
\mu\,a_{-1}}{\mu\,b_{1}^{3}}\right)
\end{gathered}\end{equation}

Solution \eqref{A2.6} satisfies Eq. \eqref{A2.1} again at
$k=\frac{\nu}{10\, \mu}$. Solution \eqref{A2.6} can be transformed
as well by means of the following set of equalities

\begin{equation}\begin{gathered}
\label{A2.7}
u=\frac{-\frac{3}{100}\left(\frac{16\nu^2\,b_{1}^{3}+225\,
\varepsilon\,\mu\,a_{-1}}{\varepsilon\,\mu\,b_{1}^{3}}\right)\,e^{2\,\eta}-
\frac{1}{100}\frac{-16\,\nu^2\,b_{1}^{3}+675\,\varepsilon\,\mu\,a_{-1}}
{\varepsilon\,\mu\,b_{1}^{2}}\,e^{\eta}+a_{-1}\,e^{-\eta}}
{e^{2\eta}+b_1\,e^{\eta}-\frac{4\,b_{1}^{3}}{27}\,e^{-\eta}}=\\
\\=
\frac{-\frac{675\,a_{-1}}{100\,b_{1}^{3}}\left(e^{\eta}+\frac{2}{3}\,b_{1}\right)^2\,
\left(e^{\eta}-\frac{b_1}{3}\right)-\frac{48\,\nu^2}{100\,\varepsilon\,\mu}
\,e^{2\,\eta}\left(e^{\eta}-\frac{b_1}{3}\right)}{\left(e^{\eta}+
\frac{2}{3}\,b_{1}\right)^2\,
\left(e^{\eta}-\frac{b_1}{3}\right)}=\\
\\
=-\frac{27\,a_{-1}}{4\,b_{1}^{3}}-\frac{12\,\nu^2}{25\,
\varepsilon\,\mu\,\left(1+\frac23\,b_1\,e^{-2\,\eta}\right)}
\end{gathered}\end{equation}

The last expression can be transformed to the exact solution
\eqref{A2.2} of Eq. \eqref{A2.1} if we assume
 $k=\frac{\nu\,z}{10\,\mu}$,
$-\frac{27\,a_{-1}}{4\,b_{1}^{3}}=C_0-\frac{6\,\nu^2}{25\,\mu}$,
$x_0=-\ln{\frac{2\,b_1}{3}}$ at $\varepsilon=1$. We do  not
new exact solitary solution of the Korteweg - de Vries - Burgers
equation again.

Two other exact solutions of work \cite{Soliman01} (expressions (31)
and (32)) do not satisfy the Korteweg - de Vries - Burgers equation
except trivial case $k=0$ and we do not study them.

We have the following results of our analysis. Using the Exp-function
method Soliman in \cite{Soliman01}
obtained exact solutions in the cumbersome form and has not simplified
these solutions. Unfortunately he has not found any new exact solutions
of the Korteweg - de Vries - Burgers equation. His statement that "all
the exact solutions of the KdV - Burgers equation are new" is not
correct.

\section{Application of the Exp-function method to the Riccati equation by Zhang}

Using the Exp-function method Zhang  presented  "new generalized
solitonary solutions of Riccati equation" in \cite{Zhang02}. Let us
illustrate that all solutions of Riccati equation are well known and
can be included in the general solution of this equation.

Author in \cite{Zhang02} has considered the Riccati equation in the
form
\begin{equation}\begin{gathered}
\label{Eq.1} \Phi_{\xi} = q +p \Phi^2
\end{gathered}\end{equation}

The general solution of Eq. \eqref{Eq.1} can be written as
\cite{Kudryashov02, Polyanin01}

\begin{equation}\begin{gathered}
\label{Eq.6} \Phi(\xi)=i\sqrt{\frac{q}{p}} \,
\frac{1+\exp(2i\sqrt{p\,q}\,(\xi-\xi_0))}{1-\exp(2i\sqrt{p\,q}\,(\xi-\xi_0))}
\end{gathered}\end{equation}

It is well known that the Riccati equation can be transformed to the
linear equation of the second order. Using transformation
\begin{equation}\begin{gathered}
\label{Eq.7} \Phi(\xi)=-\frac{1}{p}\frac{\Psi_{\xi}}{\Psi}, \quad
\Psi \equiv \Psi(\xi), \quad \Psi_{\xi}=\frac{d\Psi}{d\xi}
\end{gathered}\end{equation}

we have  the linear equation from Eq. \eqref{Eq.1}
\begin{equation}\begin{gathered}
\label{Eq.9} \Psi_{\xi\xi}+p\,q\,\Psi=0
\end{gathered}\end{equation}

General solution of Eq. \eqref{Eq.9} takes form
\begin{equation}\begin{gathered}
\label{Eq.10}
\Psi(\xi)=C_1\exp(-i\sqrt{p\,q}\,\xi)+C_2\exp(i\sqrt{p\,q}\,\xi)
\end{gathered}\end{equation}

Using \eqref{Eq.7} and \eqref{Eq.10} we get the general solution of
Eq. \eqref{Eq.9} in the form
\begin{equation}\begin{gathered}
\label{Eq.11} \Phi(\xi)=i\sqrt{\frac{q}{p}} \,
\frac{C_1\exp(-i\sqrt{p\,q}\,\xi)-C_2\exp(i\sqrt{p\,q}\,\xi)}
{C_1\exp(-i\sqrt{p\,q}\,\xi)+C_2\exp(i\sqrt{p\,q}\,\xi)}
\end{gathered}\end{equation}

Solution \eqref{Eq.11} can be transformed to solution \eqref{Eq.6}
of Eq. \eqref{Eq.1} using the following equalities
\begin{equation}\begin{gathered}
\label{Eq.12} \Phi(\xi)=i\sqrt{\frac{q}{p}} \,
\frac{C_1\exp(-i\sqrt{p\,q}\,\xi)-C_2\exp(i\sqrt{p\,q}\,\xi)}{C_1
\exp(-i\sqrt{p\,q}\,\xi)+C_2\exp(i\sqrt{p\,q}\,\xi)}=\\
\\
=i\sqrt{\frac{q}{p}} \,
\frac{1-\frac{C_2}{C_1}\exp(2i\sqrt{p\,q}\,\xi)}{1+\frac{C_2}{C_1}
\exp(2i\sqrt{p\,q}\,\xi)}= i\sqrt{\frac{q}{p}} \,
\frac{1+\exp(2i\sqrt{p\,q}\,(\xi-\xi_0))}{1-\exp(2i\sqrt{p\,q}\,(\xi-\xi_0))}
\end{gathered}\end{equation}

where
\begin{equation}\begin{gathered}
\label{Eq.13} \xi_0=\frac{i}{2\sqrt{pq}}ln
\left(-\frac{C_1}{C_2}\right)
\end{gathered}\end{equation}

It is clear that we can use general solution of Eq.
\eqref{Eq.1} taking different forms into consideration but all these forms
of solutions are the same solutions. We can write solution
\eqref{Eq.6} and \eqref{Eq.11} using hyperbolic functions,
trigonometric functions and so on. But we have to take into account
that solution \eqref{Eq.6} is the general solution and other  "new solutions"
of equation \eqref{Eq.1} can not be
found by any method.

Using the Exp-function method Zhang has found solution of Eq.
\eqref{Eq.1} in \cite{Zhang02} for the following three cases:
\begin{equation*}
1) \, p = -\frac{1}{q} ; \quad 2) \, p= \frac{1}{q} ; \quad 3) \, p=
-\frac{1}{4q}.
\end{equation*}

However all his solutions corresponds to expression \eqref{Eq.11}.
Let us demonstrate an example of reducible solution (Solution (27) in
\cite{Zhang02}) in the case $p=-\frac{1}{4q}$. Zhang has presented
solution in the form
\begin{equation}\begin{gathered}
\label{Eq.21} \Phi(\xi)= \frac{2 b_1 q
\exp(\xi+\omega)+a_0+\frac{a_0^2-4b_0^2q^2
}{8b_1q}\exp(-\xi-\omega)}{b_1\exp(\xi+\omega)+b_0-\frac{a_0^2-4b_0^2q^2
}{16b_1q^2}\exp(-\xi-\omega)}
\end{gathered}\end{equation}

However this solution can be transformed to \eqref{Eq.6}. It follows from equalities
\begin{equation}\begin{gathered}
\label{Eq.22} \Phi(\xi)=
2q\frac{16b_1^2q^2\exp(\xi+\omega)+8a_0b_1q+(a_0^2-4b_0^2q^2)
\exp(-\xi-\omega)}{16b_1^2q^2\exp(\xi+\omega)+16b_0b_1q^2-
(a_0^2-4b_0^2q^2)\exp(-\xi-\omega)}=\\
\\
=2q\frac{(a_0+4b_1q\exp(\xi+\omega))^2-4b_0^2q^2}{4q^2
(b_0+2b_1\exp(\xi+\omega))^2-a_0^2}=\\
\\
=2q\frac{(a_0+4b_1q\exp(\xi+\omega)-2b_0q)(a_0+4b_1q
\exp(\xi+\omega)+2b_0q)}{(2qb_0+4b_1q\exp(\xi+\omega)-
a_0)(2qb_0+4b_1q\exp(\xi+\omega)+a_0)}=\\
\\
=2q\frac{4b_1q\exp(\xi+\omega)+a_0-2b_0q}{4b_1q\exp(\xi+
\omega)-a_0+2b_0q}=2q\frac{C_1\exp{\xi}-C_2}{C_1\exp{\xi}+C_2}
\end{gathered}\end{equation}
where
\begin{equation}\begin{gathered}
\label{Eq.23} C_1=4b_1q\exp{\omega} ; \quad C_2=-a_0+2b_0q;
\end{gathered}\end{equation}
From the last expression we can see that solution \eqref{Eq.21} is
not "new generalized solitonary solution" of the Riccati equation.

Solution (29) in \cite{Zhang02} can be transformed as well. Zhang
have presented exact solution of the Riccati equation \eqref{Eq.1}
in the form
\begin{equation}\begin{gathered}
\label{Eq.25}
\Phi(\xi)=2q\frac{\exp{\xi}+2\sqrt{2q}+q\exp(-\xi)}{\exp{\xi}+2\sqrt{q}-q\exp(-\xi)}
\end{gathered}\end{equation}

However this solution can be simplified to solution \eqref{Eq.11} of
Eq. \eqref{Eq.1} at $p=-\frac{1}{4q}$ again by means
of equalities
\begin{equation}\begin{gathered}
\label{Eq.26} \Phi(\xi)=2q\frac{\exp(2\xi)+2\sqrt{2q}\exp{\xi}+2q-q}
{\exp(2\xi)+2\sqrt{2q}\exp{\xi}+q-2q}=\\
\\
=2q\frac{(\exp{\xi}+\sqrt{2q})^2-q}{(\exp{\xi}+\sqrt{q})^2-2q}=\\
\\
=2q\frac{(\exp{\xi}+\sqrt{2q}-\sqrt{q})(\exp{\xi}+\sqrt{2q}+
\sqrt{q})}{(\exp{\xi}+\sqrt{q}-\sqrt{2q})(\exp{\xi}+\sqrt{q}+
\sqrt{2q})}=\\
\\
=2q\frac{\exp{\xi}+\sqrt{2q}-\sqrt{q}}{\exp{\xi}-\sqrt{2q}+\sqrt{q}}=
2q\frac{\exp{\xi-C_2}}{\exp{\xi+C_2}}
\end{gathered}\end{equation}
where
\begin{equation}\begin{gathered}
\label{Eq.27} \quad C_2=\sqrt{q}+\sqrt{2q};
\end{gathered}\end{equation}

So the Exp-function method does not allow Zhang in \cite{Zhang02} to find
out any new solution
of the Riccati equation. We observe the same case as for the Burgers -
Korteweg - de Vries equation by Soliman \cite{Soliman01}.

\section{Application of the Exp-function method to
the Klein - Gordon and to the Sharma - Tasso - Olver equations by
Bekir and Boz}

Bekir and Boz in \cite{Bekir01} applied the Exp - function method to
search for exact solutions of the Klein - Gordon, the Burgers -
Fisher and the Sharma - Tasso - Olver equation.

For the Klein - Gordon equation
\begin{equation}\begin{gathered}
\label{1.3}E_1[u]= u_{tt}-u_{xx}-u+u^3=0,
\end{gathered}
\end{equation}
authors obtained the exact solutions of Eq. \eqref{1.3} in the
form (formula (3.15) in \cite{Bekir01})
\begin{equation}\begin{gathered} \label{1.4}
u(x,t)=\frac{\exp{(k\,x+\omega\,t)-\frac14\,b_{0}^{2}\,
\exp{(-(kx+\omega\,t))}}}
{\exp{(k\,x+\omega\,t)+b_0+\frac14\,b_{0}^{2}\,\exp{(-(kx+\omega\,t))}}}
\end{gathered}
\end{equation}
Solution \eqref{1.4} can be transformed into simple form if we
use the following equalities
\begin{equation}\begin{gathered} \label{1.5}
u(x,t)=\frac{\exp{(k\,x+\omega\,t)-\frac14\,b_{0}^{2}\,
\exp{(-(kx+\omega\,t))}}}
{\exp{(k\,x+\omega\,t)+b_0+\frac14\,b_{0}^{2}\,\exp{(-(kx+\omega\,t))}}}=\\
\\
=\frac{\left(\exp{(\frac12(kx+\omega\,t))}\right)^2-\left(\frac{b_0}{2}\exp{(-
\frac12(kx+\omega\,t))}\right)^2}{\left(\exp{(
\frac12(kx+\omega\,t)}+\frac{b_0}{2}\exp{(-
\frac12(kx+\omega\,t)}\right)^2}=\\
\\
=\frac{\exp{( \frac12(k\,x+\omega\,t)}-\frac{b_0}{2}\exp{(-
\frac12(kx+\omega\,t)}}{\exp{(
\frac12(kx+\omega\,t)}+\frac{b_0}{2}\exp{(- \frac12(kx+\omega\,t)}}
\end{gathered}
\end{equation}
The last solution can be easy found using the tanh-function method
\cite{Lou01, Parkes01, Malfliet01}, the simple equation method
\cite{Kudryashov03, Kudryashov04, Kudryashov05} or the singular
manifold method \cite{Weiss01, Weiss02, Kudryashov06, Kudryashov07,
Kudryashov08, Kudryashov09, Efimova01, Conte01, Conte02, Peng01}.

Bekir and Boz  studied exact solutions of the
Sharma - Tassa - Olver equation in \cite{Bekir01}
\begin{equation}\begin{gathered}
\label{03} u_t+\alpha\,\left(u^3\right)_x+\frac32\,\alpha\,
\left(u^2\right)_{xx}+\alpha\,u_{xxx}=0
\end{gathered}
\end{equation}
 Authors have looked for exact solutions of Eq. \eqref{03} using the traveling wave
\begin{equation}\begin{gathered}
\label{04} u(x,t)=u(\xi), \qquad \xi=x-w\,t
\end{gathered}
\end{equation}

Some solutions in \cite{Bekir01} do not satisfy Eq. \eqref{03}.
However exact solution (5.18) in \cite{Bekir01} in the form
\begin{equation}\begin{gathered}
\label{3.6}u(x,t)=\frac{a_0-k\,b_{-1}\,\exp{(-(k\,x+\omega\,t))}}{\exp{(k\,x+
\omega\,t)}-\frac{k^2\,b_{-1}+a_{0}^2}{k\,a_0}+b_{-1}\,\exp{(-(k\,x+\omega\,t))}}
\end{gathered}
\end{equation}
is correct but this exact solution can be
transformed taking the following equalities into account
\begin{equation}\begin{gathered}
\label{3.7}u(x,t)=\frac{a_0-k\,b_{-1}\,\exp{(-(k\,x+\omega\,t))}}{\exp{(k\,x+
\omega\,t)}-\frac{k^2\,b_{-1}+a_{0}^2}{k\,a_0}+b_{-1}\,\exp{(-(k\,x+\omega\,t))}}=\\
\\
=\frac{a_0-k\,b_{-1}\,\exp{(-(k\,x+\omega\,t))}}
{\left(\frac{1}{a_0}\,\exp{(k\,x+\omega\,t)-\frac{1}{k}}\right)
\left(a_0-k\,b_{-1}\,\exp{(-(k\,x+\omega\,t))}\right)}=\\
\\
=\frac{a_0\,k}{k\,\exp{(k\,x+\omega\,t)}-a_0}
\end{gathered}
\end{equation}

Solution
\begin{equation}\begin{gathered}
\label{3.8}u(x,t)=\frac{a_0\,k}{k\,\exp{(k\,x+\omega\,t)}-a_0}
\end{gathered}
\end{equation}
satisfy Eq. \eqref{03} but this solution can be easy found
using other methods.

More then that using the truncated expansion \cite{Kudryashov01,
Kudryashov03, Kudryashov08, Kudryashov09, Efimova01, Conte01,
Conte02, Peng01}
\begin{equation}\begin{gathered}
\label{3.1} u(x,t)=\frac{F_x}{F},\qquad F\equiv F(x,t)
\end{gathered}
\end{equation}
we can transform the Sharma - Tasso - Olver equation \eqref{03} to
the linear equation of the third order.

We obtain
\begin{equation}\begin{gathered}
\label{3.2}E_3[u]=u_t+\alpha\,\left(u^3\right)_x+\frac32\,\alpha\,
\left(u^2\right)_{xx}+\alpha\,u_{xxx}=\\
\\
=\frac{\partial}{\partial x}\left( \frac{F_t+\alpha\,F_{xxx}}{F}
\right)=0
\end{gathered}
\end{equation}
Using formula \eqref{3.1} and linear partial differential equation of the third order
\begin{equation}\begin{gathered}
\label{3.2a} F_t+\alpha\,F_{xxx}=0
\end{gathered}
\end{equation}
we can have many exact solutions of
the Sharma - Tasso - Olver equation \eqref{03}.

\section{Comparison of the Exp-function method with other methods}

We know that all nonlinear partial differential equations can be
separated on three types.

To the first type we can attribute all integrable partial differential
equations. Partial differential equations of this type have the
infinity amount of the exact solutions. The most known equations of
this type are the Korteweg - de Vries equation, the Sine - Gordon
equation, the nonlinear Schrodinger equation, the modified Korteweg
- de Vries equation, the Boussinesq equation and the Kadomtsev -
Petviashvili equation. This list can be continued but we believe that
mentioned equations are basic integrable equations. The Cauchy
problems for these equations can be solved using the inverse
scattering transform \cite{Ablowitz01, Ablowitz02, Ablowitz03}.
Solitary wave solutions can be found for these equations taking the
Hirota method into consideration \cite{Hirota01}.

The Exp-function method can be applied to these equations as well
but we do not think that the Exp-function method is better than the
Hirota method.

We can attribute the Burgers equation and other linearized
differential equations to the first type as well. One can use the
Cole - Hopf transformations \cite{Hopf01, Cole01} and other
transformations for these equations to obtain a lot of exact
solutions .

Nonlinear partial differential equations without exact solutions
belong to the second type of equations. There are a lot of
examples of such equations but we give here only simple
generalization of the Korteweg - de Vries equation in the form
\begin{equation}\begin{gathered}
\label{4.3}u_t+6\,u\,u_x+u_{xxx}+\alpha\,u=0
\end{gathered}
\end{equation}
We have not got any method to look for exact solutions of such
equations.

We can conclude all nonintegrable partial differential equations with some
exact solutions to the third type of nonlinear differential
equations. The Kuramoto - Sivashinsky equation, the Ginzburg -
Landau equation, the Korteveg - De Vries - Burgers equation, the
Fisher equation, the Fitzhugh - Nagumo equation and the Burgers -
Huxley equation are the most known equations of this type. We have many
different methods to search for exact solutions of such equations.

In last few decades great progress was made in the development
of methods for finding exact solutions of nonlinear differential
equations of the third type.

We can mention the singular manifold method \cite{Weiss01, Weiss02,
Kudryashov06, Kudryashov07, Kudryashov08, Kudryashov09, Efimova01,
Conte01, Conte02, Peng01}, tanh-function method \cite{Lou01,
Parkes01, Malfliet01} and the simple equation method
\cite{Kudryashov03, Kudryashov04, Kudryashov05, Fu01, Liu01, Peng02,
Peng03, Kudryashov10, Kudryashov11, Kudryashov12, Kudryashov13,
Fahmy01, Elgarayhi01}. Certainly the mentioned methods can be
applied to nonlinnear integrable differential equations as well but
what for? We think the Exp-function method cedes to the
enumerated methods because there are deficiencies which we
demonstrated in sections 2 - 4.

From our point of view there is no single best method to search exact
solutions of the nonlinear differential equations of the third type.
Certainly each investigator of differential equations has his
experience and his sympathy to methods but the choice of the method
depends on form of the nonlinear differential equation and the
pole of his solution.

One can think that there is the class of the nonlinear differential
equations of the third type for effective application of the
Exp-function method. This class has exact solutions with pole of
the first order. The Fitzhugh - Nagumo equation, the Burgers -
Huxley equation and some other equations \cite{Kudryashov08,
Kudryashov09} can be included to this class of equations.

However we prefer to use the singular manifold method for such
equations as well. Let us demonstrate this  approach to look for exact solutions of the
Fitzhugh - Nagumo equation \cite{Kudryashov02, Kudryashov08,
Kudryashov09, Efimova01}.

Let us take this equation in the form
\begin{equation}\label{4.5}
u_t-u_{xx}+u(1-u)(\alpha-u)=0
\end{equation}
Substituting
\begin{equation}
\label{4.6}u=\sqrt{2}\,\frac{F_x}{F}, \qquad F=F(x,t)
\end{equation}
into Eq. \eqref{4.5} and  equating
expressions at $F^{-1}$ and $F^{-2}$ to zero we have the following equations
\begin{equation}
\label{4.7}F_{xt}+\alpha\,F_{x}-F_{xxx}=0
\end{equation}
\begin{equation}
\label{4.8}F_{t}-3\,F_{xx}+\sqrt{2}\,(\alpha+1)\,F_{x}=0
\end{equation}
Solutions of this overdetermined system of equations can be easy
found. Substituting $F_t$ from \eqref{4.8} into Eq.
\eqref{4.7} we have
\begin{equation}
\label{4.9}2\,F_{xxx} - \sqrt{2}\,
(\alpha+1)\,F_{xx}+\alpha\,F_{x}=0
\end{equation}
Solution of this equation can be presented in the form
\begin{equation}
\label{4.10}F{(x,t)}=C_0(t)+C_1(t)\,e^{\lambda_1\,x}+C_2(t)\,
e^{\lambda_2\,x}, \quad \alpha\neq 1
\end{equation}
where $\lambda_{1,2}$ are nonzero roots of equation
\begin{equation}
\label{4.11}2\,\lambda^3- \sqrt{2}\,
(\alpha+1)\,\lambda^2+\alpha\,\lambda=0
\end{equation}
Solving Eq. \eqref{4.11} we have
\begin{equation}
\label{4.12}\lambda_0=0, \quad \lambda_1=\frac{\sqrt{2}}{2}, \quad
\lambda_2 = \frac{\alpha\,\sqrt{2}}{2}, \quad \alpha\neq 1
\end{equation}
and $F(x,t)$ in the form
\begin{equation}
\label{4.13}F{(x,t)}=C_0(t)+C_1(t)\,e^{{x\,\sqrt{2}}/{2}}+C_2(t)\,
e^{{\alpha\,x\,\sqrt{2}}/{2}}
\end{equation}

Functions $C_0(t)$, $C_1(t)$  and $C_2(t)$ can be obtained after
substitution \eqref{4.13} into Eq. \eqref{4.8}. We have
\begin{equation}\begin{gathered}
\label{4.14}F{(x,t)}=c_0+c_1\,e^{(x\,\sqrt{2}/{2}- \alpha\,t+t/2)}
+c_2\, e^{\,(\alpha\,x\sqrt{2}/2-\alpha\,t+\alpha^2\,t/2\,)}, \qquad
\alpha\neq 1
\end{gathered}\end{equation} where $c_0$,
$c_1$ and $c_2$ are arbitrary constants.

Substituting solutions for $F(x,t)$ into formula \eqref{4.6} for $u$
we have exact solutions of the Fitzhugh - Nagumo equation in the
form
\begin{equation}\begin{gathered}
\label{4.15}u=\frac{c_1\,e^{(x\,\sqrt{2}/{2}- \alpha\,t+t/2)}
+c_2\,\alpha\,
e^{\alpha\,/2\,\left(x\sqrt{2}+\alpha\,t-2\,t\,\right)}}
{c_0+c_1\,e^{(x\,\sqrt{2}/{2}- \alpha\,t+t/2)} +c_2\,
e^{\alpha\,/2\,\left(x\sqrt{2}+\alpha\,t-2\,t\,\right)}}, \qquad
\alpha\neq 1
\end{gathered}\end{equation}

At $\alpha=1$  we can obtain $F(x,t)$ from the set of Eqs.
\eqref{4.7} - \eqref{4.6} as well. It takes the form

\begin{equation}\begin{gathered}
\label{4.16}F{(x,t)}=c_0+(c_1 +c_2\,
x+\sqrt{2}\,c_2\,t)\,e^{\,(x\sqrt{2}/2-\,t/2\,)},
\end{gathered}\end{equation}
Exact solution can be found from formula \eqref{4.6}
\begin{equation}\begin{gathered}
\label{4.17}u=\frac{\sqrt{2}\,(c_1\,\sqrt{2}+2\,c_2+c_2\,\sqrt{2}\,x+2\,c_2\,t)\,
e^{\,(x\sqrt{2}/2-\,t/2\,)}} {2\,\left(c_0+(c_1 +c_2\,
x+\sqrt{2}\,c_2\,t)\,e^{\,(x\sqrt{2}/2-\,t/2\,)}\right)},
\end{gathered}\end{equation}

The last solution can not be found by application of the
Exp-function method.  Exact solution \eqref{4.15} was not found by
means of the Exp-function method as well but it could be found.
However we think the application of the singular manifold approach
for constructing exact solution to the Fitzhugh - Nagumo equation
is easier than the application of the Exp-function method.

\section{Modified simple equation method}

The simple equation method is applied to find out an exact solution of
a nonlinear ordinary differential equation
\begin{equation}\label{ODE}
P(y, y', y'', y''', \dots) = 0,
\end{equation}
where $y=y(z)$ is an unknown function, $P$ is a polynomial of the
variable $y$ and its derivatives.

To solve Eq.~(\ref{ODE}) we expand its solutions $y(z)$ in a finite
series
\begin{equation}\label{expansion}
y(z) = \sum_{k=0}^N A_k Y^k, \quad A_k = \textrm{const}, \quad A_N
\neq 0,
\end{equation}
where $Y=Y(z)$ are some special functions. These are, for example,
the functions $\tanh(kz)$ for the tanh--method.

The basic idea of the simple equation method is the assumption
that $Y=Y(z)$ are not only some special functions, but they are the
functions that satisfy some ordinary differential equations. These
ordinary differential equations are referred to as the simplest
equations. Two main features characterize the simplest equation:
first, this is the equation of a lesser order than Eq.~(\ref{ODE});
second, the general solution of this equation is known (or we know
the way of finding its general solution). This means that the exact
solutions $y(z)$ of Eq.~(\ref{ODE}) can be presented by a finite
series (\ref{expansion}) in the general solution $Y=Y(z)$ of the
simplest equation.

One of the simple equation is the Riccati equation
\begin{equation}\label{Riccati}
Y' + Y^2 + aY + b = 0, \quad a, b = \textrm{const}.
\end{equation}

The general solution of Eq.~(\ref{Riccati}) is usually searched with
the help of the standard anzats \cite{Kudryashov02, Polyanin01}
\begin{equation}\label{anzats}
Y = \psps,
\end{equation}
where $\psi = \psi(z)$ is an unknown function to be found. This
anzats leads to the second order linear ordinary differential
equation
\begin{equation}\label{LODE_II}
\psi'' + a\psi' + b\psi = 0
\end{equation}
that general solution $\psi=\psi(z)$ is well--known. Turning back to
the variable $Y$ (\ref{anzats}) by applying the general solution
$\psi=\psi(z)$ and substituting the ratio (\ref{anzats}) in the
expansion (\ref{expansion}) we immediately obtain the exact solution
$y=y(z)$ of Eq.~(\ref{ODE}).

Meanwhile, if one feels oneself ill at ease with differential
equations then one can exclaim: How can I find this magic simplest
equation? To get over this threatening obstacle we suggest to avoid
this difficulty by writing the expansion (\ref{expansion}) straight
in the form
\begin{equation}
y(z) = \sum_{k=0}^N A_k Y^k = \sum_{k=0}^N A_k \left( \psps
\right)^k.
\end{equation}

Therefore, the exact solutions $y=y(z)$ of the nonlinear ordinary
differential equation~(\ref{ODE}) we could look for in the form
\begin{equation}\label{expansion_psi}
y(z) = \sum_{k=0}^N A_k \left( \psps \right)^k, \quad A_k =
\textrm{const}, \quad A_N \neq 0,
\end{equation}
where the function $\psi=\psi(z)$ obeys Eq.~(\ref{LODE_II}).

Another simple equations are discussed in~\cite{Kudryashov07}.

In the present paper we extend the simple equation method by the
assumption that the function $\psi=\psi(z)$ is the general solution
for the linear ordinary differential equation of the third order
\begin{equation}\label{LODE_III}
\psi''' = \alpha\psi'' + \beta\psi' + \gamma\psi, \quad \alpha,
\beta, \gamma = \textrm{const}.
\end{equation}

Now we are going to find the exact solutions of some equations like
Eq.~(\ref{ODE}) by using the extended simple equation method. This
implies that we will search the solution $y=y(z)$ of Eq.~(\ref{ODE})
in a form of the expansion (\ref{expansion_psi}), where the function
$\psi=\psi(z)$ obeys Eq.~(\ref{LODE_III}) and the coefficients $A_k$
and the parameters $\alpha$, $\beta$, $\gamma$ are to be found.

Below we look round the main steps of our algorithm.

Firstly, for Eq.~(\ref{ODE}) we determine the positive number $N$ in
the expansion (\ref{expansion_psi}). To realise this procedure we
concentrate our attention on the leading terms of Eq.~(\ref{ODE}).
These are the terms that lead to the least positive $p$ when
 a monomial $y=\frac{a}{z^p}$ is substituted in all the items of this
equation. The homogeneous balance between the leading terms provides
us the value of $N$. This value is also referred to
the order of a pole for the solution of Eq.~(\ref{ODE}).

Secondly, we substitute in Eq.~(\ref{ODE}) the expansion
(\ref{expansion_psi}) with the value of $N$ already determined, we
calculate all the necessary derivatives $y'$, $y''$, $y'''$, $\dots$
of an unknown function $y=y(z)$ and we account that the function
$\psi=\psi(z)$ satisfies Eq.~(\ref{LODE_III}). As a result of this
substitution we get a polynomial with respect to the ratio $\psps$
and its derivative $\Dpsps$.

Thirdly, in the polynomial just obtained we gather the items with the
same powers of the ratio $\psps$ and its derivative $\Dpsps$ and we
equate with zero all the coefficients of this polynomial. This
operation yields a system of algebraic equations with respect to the
coefficients $A_k$ of the expansion (\ref{expansion_psi}) and to the
parameters $\alpha$, $\beta$, $\gamma$ of Eq.~(\ref{LODE_III}).

Fourthly, we solve the algebraic system.

Fifthly, in Eq.~(\ref{LODE_III}) we take the parameters $\alpha$,
$\beta$, $\gamma$ that are the solutions of the algebraic system and
 derive the general solution
$\psi=\psi(z)$ of Eq.~(\ref{LODE_III}) for them.

And finally, we substitute the general solution $\psi(z)$, its
derivative $\psi'(z)$ and coefficients $A_k$ in the expansion
(\ref{expansion_psi}). The expansion (\ref{expansion_psi}) written
in such form gives the exact solution of Eq.~(\ref{ODE}).

\section{Application of the modified simplest equation method to generalization
of the Korteweg - de Vries equation}

Let us apply the modified simplest equation method to look for exact solutions of
the generalized Korteweg--de~Vries equation with source in the form
\begin{equation}\label{KdV}
u_t + u_{xxx} - u_{xx} - u u_x + 3\left( u_x \right) ^2 + 3u u_{xx}
+ 3u^2 u_x + h u - 2u^2 + u^3 = 0,
\end{equation}
where $u=u(x,t)$ is an unknown function, $u_t$, $u_x$, $\dots$ are
the partial derivatives of $u(x,t)$ and $h$ is an arbitrary
constant.

Using the travelling wave let us find exact solutions of equation
\begin{equation}\label{KdV_Wave}
y''' + 3yy'' - y'' - yy' + 3\left( y' \right)^2 - C_0 y' + 3y^2 y' +
y^3 - 2y^2 + hy = 0.
\end{equation}

We look for solutions of Eq. \eqref{KdV_Wave} in the form
\begin{equation}\begin{gathered}
\label{7.3}y(z)=A_0+A_1\,\frac{\psi_z}{\psi}
\end{gathered}
\end{equation}
where $\psi(z)$ satisfies Eq. \eqref{LODE_III}. Substituting
\eqref{7.3} into Eq. \eqref{KdV_Wave} and taking Eq.
\eqref{LODE_III} into account
 we obtain
\begin{equation}\begin{gathered}
\label{7.4}A_1=1,\\
\\
A_0=\frac23-\frac{\alpha}{3},\\
\\
\beta=\frac43-\frac{{\alpha}^{2}}{3}\,-h,\\
\\
C_0=2-h,\\
\\
\delta=\frac{{\alpha}^{3}}{27}\,+{\frac
{16}{27}}-\frac{4\,\alpha}{9}\,+\frac{h\alpha}{3}\,-\frac{2\,h}{3}\,
\end{gathered}
\end{equation}
We have solution
\begin{equation}\begin{gathered}
\label{7.5}y(z)=\frac23-\frac{\alpha}{3}\,+\,\frac{\psi_z}{\psi}
\end{gathered}
\end{equation}
Where $\psi(z)$ satisfy linear equation in the form
\begin{equation}\begin{gathered}
\label{7.7}\psi_{{{zzz}}}-\alpha\psi_{{{zz}}}+ \left(
h+\frac{{\alpha}^{2}}{3}-\frac43  \right)\psi_{{z}} +\left(
\frac{2h}{3}+\frac{4\alpha}{9}-\frac{h\alpha}{3}- \frac{{\alpha
}^{3}}{27}-{\frac {16}{27}} \right)\psi=0
\end{gathered}
\end{equation}
Solution $\psi(z)$ at $h<1$ can be written in the form
\begin{equation}\begin{gathered}
\label{7.8}\psi(z)=C_{{1}}{e^{ \left(\frac{\alpha}{3}-\frac23\,
\right) z}}+C_{{2}}{e^{ \left( \frac{\alpha}{3}+\frac13+\sqrt {1-h}
\right) z}}+C_{{3}}{e^{ \left( \frac{\alpha}{3}\,+\frac13-\sqrt
{1-h} \right) z}}
\end{gathered}
\end{equation}
We have solution $y(z)$ of Eq. \eqref{KdV_Wave} at $C_0=2-h$
\begin{equation}\begin{gathered}
\label{7.9}y(z)={\frac {C_{{2}}\,\left(1+\sqrt{1-h}\right){e^{z
\left( 1+\sqrt{1-h} \right)
}}+C_{{3}}\left(1-\sqrt{1-h}\right){e^{-z
 \left( -1+\sqrt {1-h} \right)}}}{C_{{1}}+C_{{2}}{e^{z \left( 1+\sqrt {1-h} \right)
}}+C_{{ 3}}{e^{-z \left( -1+\sqrt {1-h} \right) }}}}
\end{gathered}
\end{equation}

Assuming $h=1$ from Eq. \eqref{7.7} we have solution $\psi(z)$ in
the form
\begin{equation}\begin{gathered}
\label{7.10}\psi(z)=C_{{1}}{\exp{ \left(
\frac{\alpha\,z}{3}\,-\frac{2\,z}{3} \right)
}}+\left(C_2+C_{{3}}\,z\right)\,{\exp{ \left(\frac{\alpha\,z}{3}+
\frac{z}{3}\right)}}
\end{gathered}
\end{equation}
and solution of Eq. \eqref{KdV_Wave}
\begin{equation}\begin{gathered}
\label{7.11}y(z)={\frac
{C_{{2}}+C_{{3}}(1+z)}{C_{{1}}\,{\exp{\left(-z\right)}}+C_{{2}}+C_{{3}}z}}
\end{gathered}
\end{equation}

Assuming $h=1+k^2$ in the case $h>1$ we have solution $\psi(z)$
\begin{equation}\begin{gathered}
\label{7.12}\psi(z)=C_{{1}}{e^{ \left(
\frac{\alpha\,z}{3}\,-\frac{2\,z}{3} \right)}}+C_{{2}}{e^{ \left(
\frac{\alpha\,z}{3}+\frac{z}{3}\,\right)}}\sin \left( k\,z \right)
+C_{{3}}{e^{\left(\frac{\alpha\,z}{3}+\frac{z}{3}\right)}}\cos
\left( kz \right)
\end{gathered}
\end{equation}
Exact solutions of Eq. \eqref{KdV_Wave} in this case takes the
form
\begin{equation}\begin{gathered}
y(z)={\frac {\left( C_{{2}}\,k+C_{{3}} \right) \cos \left( k\,z
\right)+ \left( C_{{2}}-C_{{3}}\,k \right) \sin \left( k\,z \right)
  }{C_{{1}}\,{\exp{\left(-
z\right)}}+C_{{2}}\sin \left( k\,z \right) +C_{{3}}\cos \left( k\,z
\right) }}
\end{gathered}
\end{equation}

These solutions were not found by means of the Exp-function method.

\section{Conclusion}

We have given the analysis of the application of the Exp-function
method for finding exact solutions of nonlinear differential
equations. On the examples of papers \cite{Soliman01, Zhang02,
Bekir01} we have shown that this method allows us to search for
exact solutions. However these exact solutions are cumbersome and as
a rule we need to simplify them. Without simplifications one can
think that we obtain "new solutions" of nonlinear differential equations.

We have discussed different methods for finding exact solutions. From our
point of view we do not have the single best method to search for exact solutions
of nonlinear nonintegrable differential equations. Sometimes we have to
apply the singular manifold method \cite{Weiss01, Weiss02, Kudryashov06,
Kudryashov07, Kudryashov08, Kudryashov09, Efimova01, Conte01, Conte02, Peng01},
tanh-function method \cite{Lou01, Parkes01, Malfliet01}, the
simple equation method \cite{Kudryashov03, Kudryashov04, Kudryashov05},
trial function method \cite{Xie01, Xie02} and so on.

In this paper we have presented the modified simple equation method and
we think this method can be used to look for exact solutions in a
number cases. We have illustrated our method to obtain exact solutions of
the generalized Korteveg - de Vries equation with source.

\section {Acknowledgements}

This work was supported by the International Science and Technology
Center under Project B 1213.

\end{document}